\input{aipcheck}

\documentclass[
    ,final            
  ]
  {aipproc}
\usepackage{graphicx}
\pdfoutput=1
\usepackage{epstopdf}
\usepackage{amsmath}
\usepackage{amssymb}
\usepackage{subfig}
\usepackage[outercaption]{sidecap}    
    \sidecaptionvpos{figure}{c} 
\usepackage{indentfirst}
\usepackage[labelsep=period, labelfont=bf]{caption}
\usepackage{dcolumn}
\usepackage{bm}
\layoutstyle{8x11single}

\begin{document}

\title{Evolution of shock instability in granular gases with viscoelastic collisions}


\author{Nick Sirmas}{
  address={Department of Mechanical Engineering, University of Ottawa, ON, K1N6N5, Canada}
}
\author{Matei Radulescu}{
}

\begin{abstract}
Shocks in granular media have been shown to develop instabilities. We address the role that early stages of shock development have on this type of instability. We look at the evolution of shock waves driven by a piston in a dilute system of smooth inelastic disks, using both discrete particle and continuum modelling. To mimic a realistic granular gas, viscoelastic collisions are approximated with an impact velocity threshold $u^*$ needed for inelastic collisions to occur. We show that behaviour of the shock evolution is dependent on the ratio of piston velocity to impact velocity threshold $u_p/u^*$, and the coefficient of restitution $\varepsilon$. For $u_p/u^*=2.0$, we recover shock evolution behaving similar to that observed in purely inelastic media. This is characterized by a short period where the shock front pulls towards the piston before attaining a developed structure. No pullback is seen for $u_p/u^*=1.0$.  Results show the onset of instability for these stronger shocks during this evolving stage. These results suggest that the early stages of shock evolution play an important role in the shock instability. 

\end{abstract}

\maketitle


\section{Introduction}
\vspace{-1pt}
\noindent Experiments have shown that shocks in granular media can become unstable. For example, unstable formations have been observed in granular media dispersed by a shock wave~\cite{Frost2012, Rodriguez2013} and for blast waves through dilute granular flows~\cite{Boudet2013}. Similar pattern formations can be seen when granular media is subjected to a vertically oscillating bed, both experimentally, and numerically~\cite{Bizon1998, Carrilo2008}. Such instabilities also develop in piston driven shock waves through a system of dilute inelastic disks (2D), with collisions of disks modeled deterministically~\cite{RadulescuSirmas2011} and via continuum modelling of Euler equations in granular media~\cite{Sirmasetal2013}.     

\begin{figure}[t]
\centering
\captionsetup{type=figure}

\subfloat[Comparison of developed shock structures for media governed by purely inelastic and viscoelastic collisions]{\includegraphics[trim=0cm 0cm 0cm .5cm, clip=true, width=0.51\linewidth]{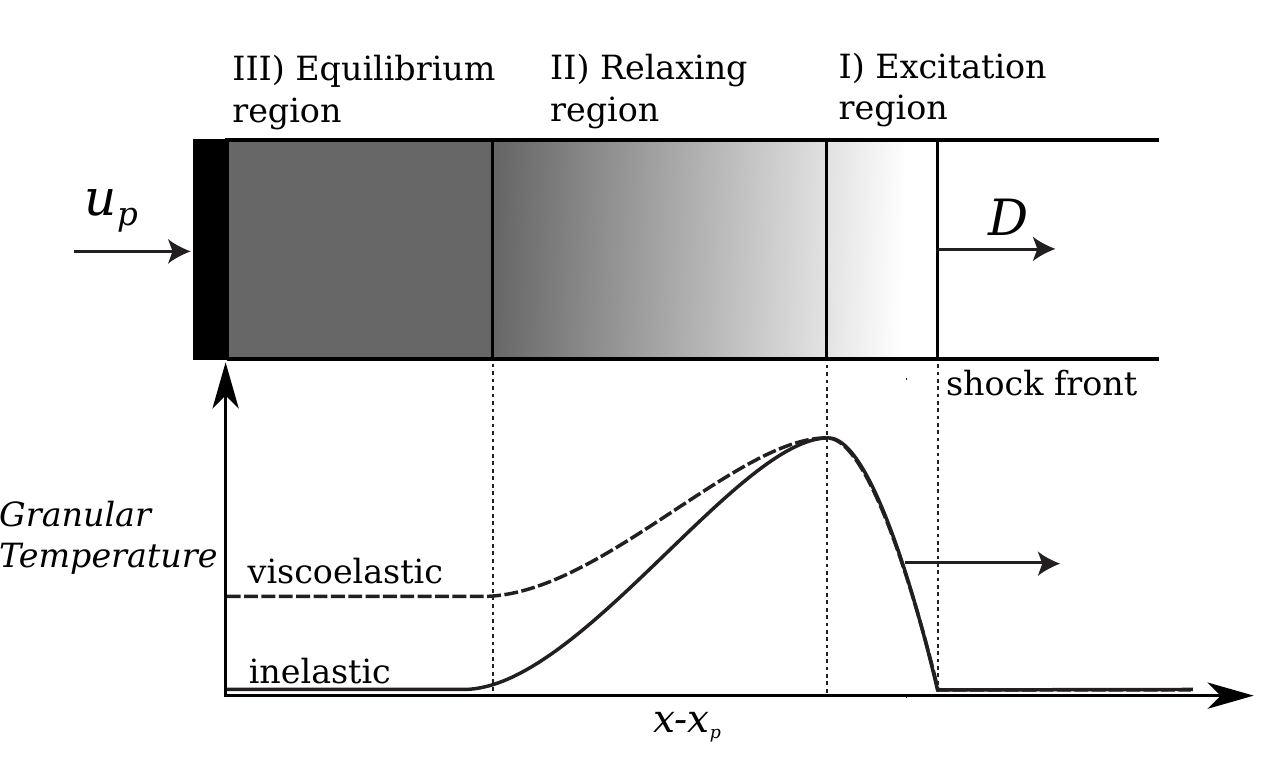}}\hspace{30pt}

\subfloat[Evolution of structure through inelastic granular media]{\includegraphics[trim=0cm 0cm 0cm 0cm, clip=true, width=0.44\linewidth]{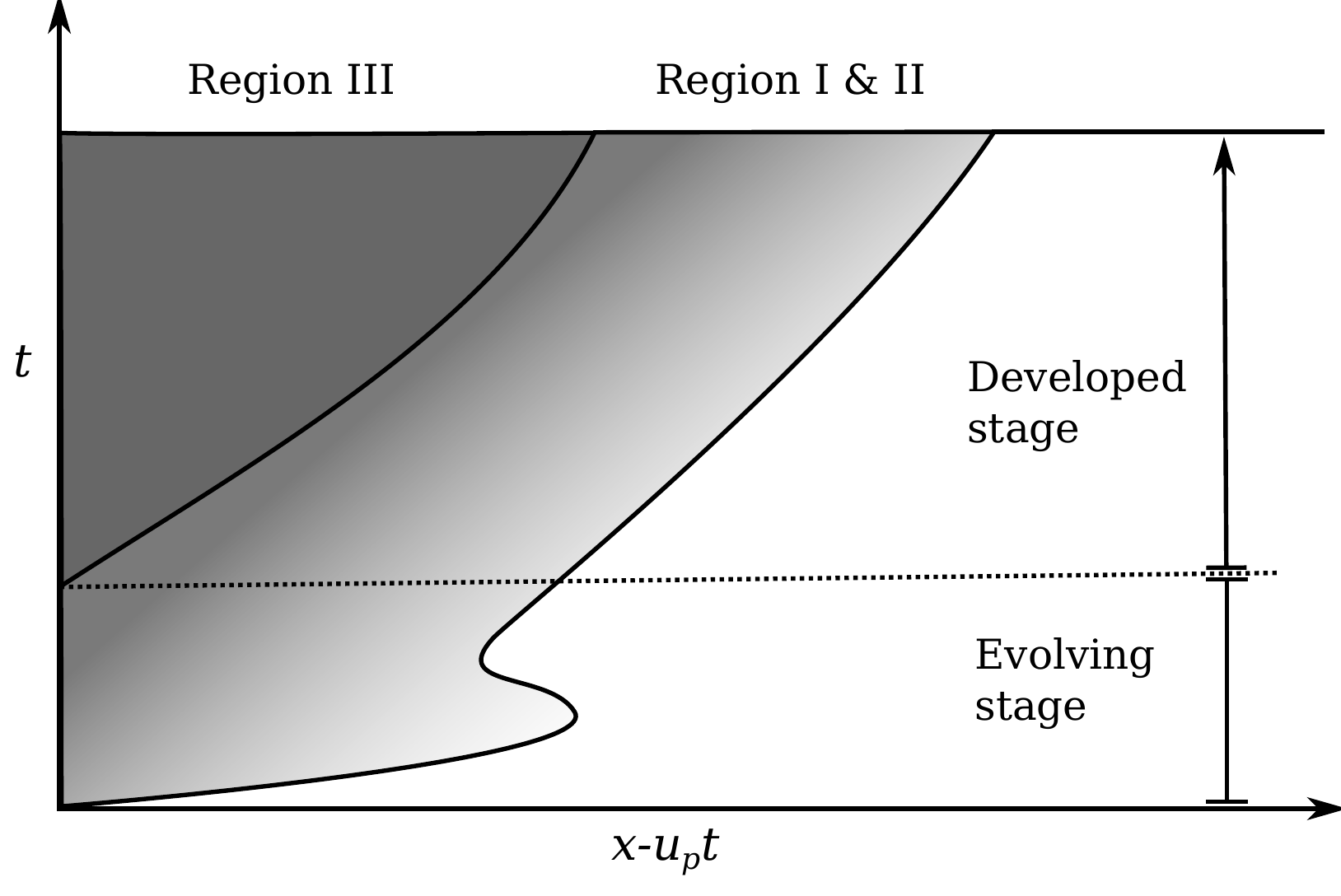}}
\caption{Sketches demonstrating structure and development of piston driven shock waves through granular media}
\label{fig:sketch}
\end{figure}

The evolution of shocks through such a medium are not well understood, and the role that the early stages have on instability is unknown. Previous investigations have looked at the unique one-dimensional structure and evolution of shock waves through granular media, although instabilities had not been identified \cite{Goldshteinetalch31996, Kamenetsky_etal2000}. Such structures have been studied via the problem of piston driven shock waves, as sketched out in Figure~\ref{fig:sketch}(a). In such a structure, a piston causes a shock wave to traverse through a granular media, which causes the granular temperature to increase (region I). Due to the inelasticity and increased rate of the collisions within this excited region, the granular temperature decreases, and density increases within the `relaxing' region (region II). Eventually, the density is high enough that the collisions subside, characterized as the `equilibrium' region (region III). When all collisions are inelastic, the equilibrium region tends to zero granular temperature. Structures in granular media, characterized by viscoelastic collisions, reveal a similar structure, although kinetic energy is retained in the equilibrium region \cite{Sirmas}.

The evolution of such a structure has previously been investigated numerically by solving the one-dimensional Euler equations through granular media~\cite{Kamenetsky_etal2000}. These studies investigate how the initial packing fraction of particles $\eta$ and the coefficient of restitution $\varepsilon$ affect the evolution. The authors also introduce an upper threshold in density whereby collisions return to being elastic, thus fixing the equilibrium state. A sketch of this evolution can be seen in Figure~\ref{fig:sketch}(b), which looks at the development of each region in an $x$ vs.~$t$ plot in the piston frame of reference. The studies conclude two distinct stages during the development of the granular shock waves. The first stage is the evolving stage, whereby the media is initially compressed and relaxation begins. Eventually, the equilibrium region is formed and the developed structure seen in Figure~\ref{fig:sketch}(a) is obtained. The most interesting behaviour is observed during the evolving stage of the shock structure. Initially, a fast travelling shock wave is produced. In the presence of all collisions being inelastic, the relaxation slows down the shock front and even pulls back towards the piston. The reason for this pullback was not investigated. Eventually, as the equilibrium region is developed, the shock front re-accelerates and tends to the developed structure, attaining constant velocity. 

The dynamics of this evolving stage are not well understood, and the role this stage has on the stability of the shock waves has not been investigated. Furthermore, this evolution through a system of particles undergoing viscoelastic collisions has not been explored. The goal of our present study is to clarify the role of this initial transient on the instabilities observed in piston driven shock waves in granular media.

Our previous studies looked at the role viscoelastic collisions have on the stability of shock waves, in order to mimic a more realistic granular media~\cite{RadulescuSirmas2011, Sirmasetal2013}. This is done by introducing an impact velocity threshold for which disks collide inelastically (a simple model for viscoelastic collisions~\cite{Poscheletal2003}). This treatment yields the formation of high density non-uniformities and convective rolls within the structure for sufficiently strong shocks. This transition to unstable structures occurs when the driving piston velocity is greater than the impact threshold. 

In the present paper we extend this work to look at the evolution of these shock waves, with inelastic collisions moderated by an impact threshold, thus allowing us to compare with the evolution observed in purely inelastic media, as well as the departure from such behaviour. This is done using both discrete particle and continuum modelling. To better understand the shock dynamics, the family of characteristics are reconstructed from the evolving shock structures. 

This paper is organized as follows. First, we describe the model that is used to study the evolution and stability of shock waves through  granular gases characterized by viscoelastic collisions. Secondly, we summarize the numerical methods we use for the molecular and continuum models. Specifics are outlined for the modifications made for the viscoelastic treatment. Finally, the evolution of the shock waves are presented, comparing results obtained using both models.  
\vspace{-1pt}
\section{Problem Overview}
\vspace{-1pt}
\noindent The medium we investigate is a system of colliding hard disks (in 2D). Each binary collision is elastic, unless the impact velocity (normal relative velocity) exceeds a velocity threshold $u^*$, a simple treatment for viscoelastic media. If the collision is inelastic, the disks collide with a constant coefficient of restitution $\varepsilon$.

The system we study is a classical shock propagation problem, whereby the motion of a suddenly accelerated piston driven into a thermalized medium drives a strong shock wave. The driving piston is initially at rest and suddenly acquires a constant velocity $u_p$. Our previous study concluded that the structure depends on the ratio $u_p/u^*$ and $\varepsilon$ (for constant initial packing factor $\eta_1$)~\cite{Sirmas}. Therefore, we maintain $u^*=10$ and $\eta_1=0.012$ throughout this investigation.

\vspace{-1pt}
\section{Numerical Models}
\vspace{-1pt}
\subsection{Molecular dynamics model}
\vspace{-1pt}
\noindent The molecular dynamics (MD) simulations are calculated using the Event Driven Molecular Dynamics technique by employing the implementation of P\"{o}schel and Schwager \cite{Poschel&Schwager2005}, that we extended to treat a moving wall (piston). The disks were initialized with equal speed and random directions.  The system was let to thermalize and attain Maxwell-Boltzmann statistics.  Once thermalized, the piston started moving with constant speed. 

The initial packing fraction of the disks was chosen to be $\eta_1=(N\pi d^2)/4A=0.012$, where $N\pi d^2/4$ is the specific volume (area) occupied by $N$ hard disks with diameter $d$, and domain area $A=L_x\times L_y$. 

All velocities and distances are normalized by the initial root mean squared velocity $u_{rms_1}$ and the initial mean free path $\lambda_1$ of the media, respectively. Where the initial mean free path can be written as $\lambda_1=1/\left(\sqrt{2\pi} d g_2 N/A\right)$ and $g_2(\eta)=(1-(7/16)\eta)((1-\eta)^2)$ is the pair correlation function for a system of hard disks \cite{Torquato1995}. Thus fixing the time scaling by ${\lambda_1}/{u_{rms_1}}$. The initial domain size is $A=172.9\lambda_1\times17.2\lambda_1$,  occupied by $N$=30,000 hard disks.

One-dimensional distributions of macroscopic properties are obtained by coarse grain averaging strips parallel to the piston with a width of $\Delta x=0.5\lambda_1$, as well as ensemble averaging over 25 simulations.
\vspace{-1pt}
\subsection{Continuum model}
\vspace{-1pt}
\noindent In the continuum model, we consider a two-dimensional granular gas, by modelling a system of smooth inelastic disks. For such a system, the Euler hydrodynamic equations for mass, momentum and energy take the form:
\begin{eqnarray}
\frac{\partial \rho}{\partial t}+\vec{\nabla}\cdot\left(\rho\vec{u}\right)&=&0\notag\\
\frac{\partial \rho\vec{u}}{\partial t}+\vec{\nabla}\cdot\left(\rho\vec{u}\vec{u}\right)&=&-{\nabla}p\\
\frac{\partial E}{\partial t}+\vec{\nabla}\cdot\left(\vec{u}(E+p)\right)&=&\zeta\notag
\end{eqnarray}
where $E=\rho T+\frac{1}{2}\rho \vec{u}\cdot\vec{u}$ is the total energy density, in terms of density $\rho$, granular temperature $T$ and velocity $\vec{u}$.
For granular systems, the hydrostatic pressure $p$ can be approximated as \cite{Brilliantov&Poschel2004}:
\begin{equation}\label{EOS1}
p=\rho e\left[ 1+(1+\varepsilon)\eta g_2(\eta)\right]
\end{equation}
where $e=T$ is the internal energy in 2D. 
Note that the equation of state given by Eq.~\eqref{EOS1} implicitly assumes that all collisions are inelastic. Since the non-activated collisions are elastic, Eq.~\eqref{EOS1} will underpredict the pressure, notably in the early stages of shock evolution, and in the equilibrated region. These discrepancies are discussed below. 

The cooling coefficient $\zeta$ for constant $\varepsilon$ may be written as \cite{Brilliantov&Poschel2004}:
\begin{equation}\label{zeta_nocorrect}
\zeta=-\frac{4}{d\sqrt{\pi}}\left(1-\varepsilon^2\right)\rho T^{3/2}\eta g_2(\eta)
\end{equation}

In our viscoelastic model, inelastic collisions occur for a fraction of collisions, making the cooling rate from \eqref{zeta_nocorrect} invalid. Therefore, a modification to the cooling rate is needed to solely account for the energy losses attributed to inelastic collisions. These modifications are done using traditional kinetic theory methods looking at binary collisions, yielding~\cite{Sirmasetal2013}:

\begin{equation}
\zeta^*=-\frac{4}{d\sqrt{\pi}}\left(1-\varepsilon^2\right)\rho T^{3/2}\eta g_2(\eta)\exp \left\{ \frac{1}{2}\frac{{u^*}^2}{u_{rms}^2}\right\}\left(1+\frac{1}{2}\frac{{u^*}^2}{u_{rms}^2}\right)\\
\end{equation}


 The software package \textit{MG} was used to solve the governing equations of the described system, courtesy of Sam Falle (University of Leeds). The \textit{MG} package utilizes a second order Godunov solver with adaptive mesh refinement. The hydrodynamics of the granular gas are solved by investigating the flow in the piston frame of reference. A reflective wall boundary was implemented on the left (piston face) with flow travelling towards the wall at constant velocity $u_p$. The upper and lower boundaries have reflective wall boundary conditions, and the right boundary has free boundary conditions. In order to compare with the MD results a domain height of $17.2\lambda$ was used. A resolution of $\Delta x=d\approx0.038\lambda_1$ was used for the one-dimensional results, and $\Delta x=\Delta y=0.343\lambda_1$ for two-dimensional results.

In order to investigate the unstable shock structure for the two-dimensional modelling, the initial and incoming flow density were perturbed. Random density perturbations with variance of $10\%$ were applied in cells with area $0.5\lambda\times 0.5\lambda$ (approximately 40 cells high in the current domain). This level of perturbation is comparable with the thermal noise at these scales, and the dimensions are found to be sufficient enough to avoid the frequency of instability of being an artifact of the cell size. 

\vspace{-1pt}
\subsection{Details for characteristics}
\vspace{-1pt}
\noindent To investigate the dynamics of the shock waves, the family of characteristics were constructed. The particles paths ($P$), forward ($C^+$) and backward ($C^-$) running characteristics on an $x$ vs. $t$ plane are given by:
\begin{equation}\label{eq:char}
P:~~\frac{dx_p}{dt}=u~~~~~~~~~~~~C^+:~~\frac{dx_+}{dt}=u+c~~~~~~~~~~~~C^-:~~\frac{dx_-}{dt}=u-c
\end{equation}
where $u$ is the local particle velocity normal to the piston and $c$ is the local speed of sound, at a given time. They represent the trajectories of fluid particles, right running pressure waves and left running pressure waves, respectively~\cite{Landau&Lifshitz1987}. The scaled speed of sound for such a media is approximated for an elastic system of disks, taken as~\cite{Sirmasetal2012}:
\begin{equation}
\frac{c}{u_{rms_1}}=\sqrt{\frac{T}{T_1}(1+(1-\eta)^{-2}+2\eta(1-\eta)^{-1})}
\end{equation}
The local packing factor is taken from the density jump, $\eta=\eta_1 {\rho}/{\rho_1}$.


The trajectories of the characteristics were obtained numerically by integrating \eqref{eq:char}. The $C^+$ characteristics are initiated from the piston face at specified intervals in time, while $C^-$ characteristics are initiated from the shock front at similar time intervals.  The particle paths are initialized at specified locations away from the initial piston position, denoted as $\xi=x(t=0)$ for each particle path.
\vspace{-1pt}
\section{Results and Discussion}
\begin{figure}[b]
	\captionsetup{type=figure}
	\subfloat[MD evolution]{\includegraphics[trim=.5cm .5cm 0cm 1cm, clip=true, width=0.51\linewidth]{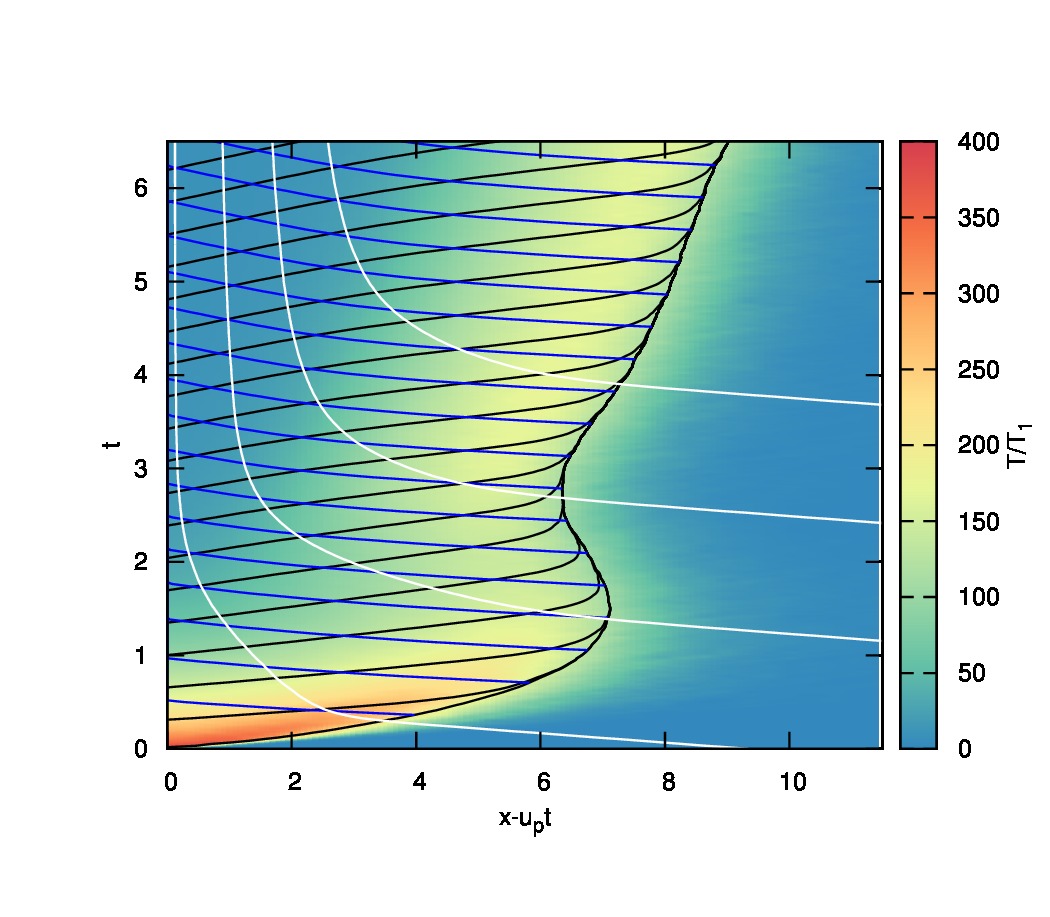}}~
	\subfloat[Continuum evolution ]{\includegraphics[trim=.5cm .5cm 0cm 1cm, clip=true, width=0.51\linewidth]{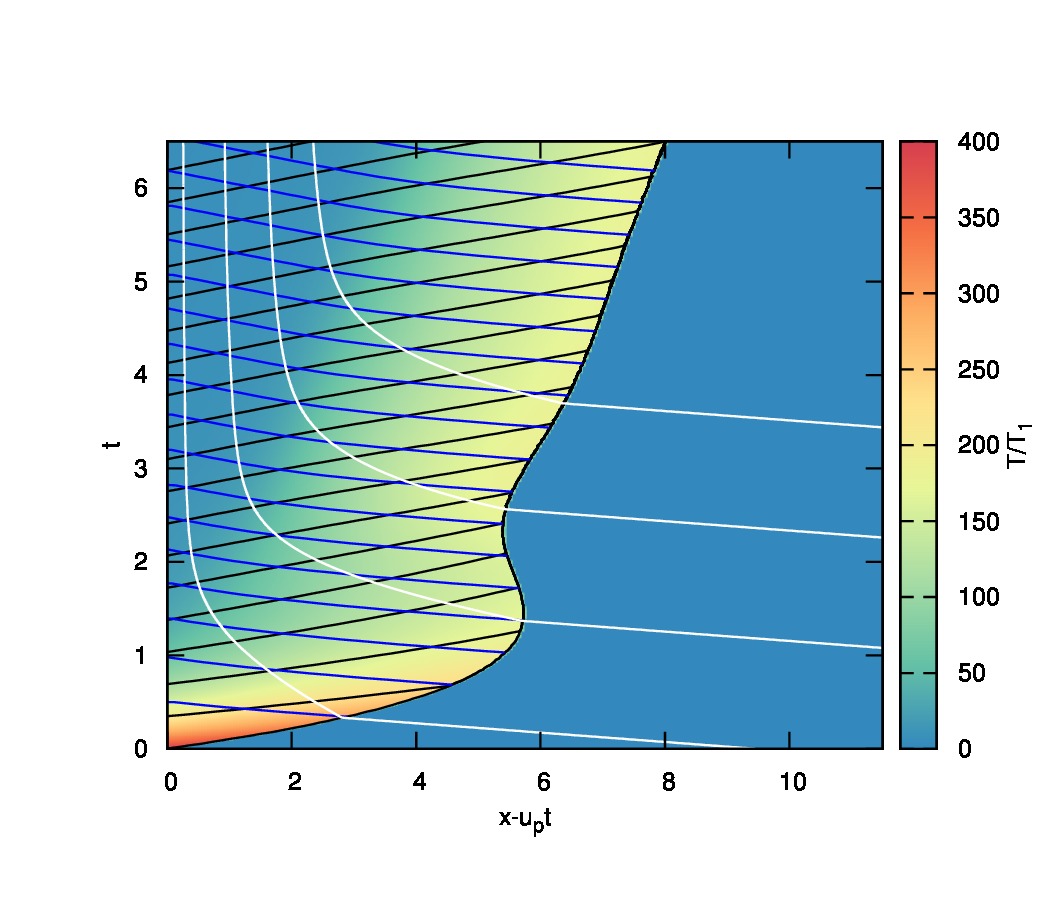}}
	\caption{Evolution of temperature distribution on an $x$ vs.~$t$ plane, in the piston frame of reference, comparing MD (left) and continuum (right) models for $u_p=20$, $u^*=10$ and $\varepsilon=0.95$. Also shown are select particle paths (white), forward (black) and backward (blue) running characteristics.}
	\label{fig:xt_pframe}
\end{figure}

\vspace{-1pt}
\noindent First we investigate the evolution of a strong shock, which has been shown to develop instabilities and whose structure behaves similar to purely inelastic granular gases. Figures \ref{fig:xt_pframe}(a) and (b) show the evolution of temperature obtained for the MD and continuum models, respectively, for $u_p/u^*=2.0$ and $\varepsilon=0.95$. The evolution is shown in the piston frame of reference, with selected particle paths, $C^+$ characteristics extending from the piston, and $C^-$ from the shock front. As expected, the $C^+$ characteristics collapse along the shock front. Since the continuum solution does not account for any viscosity, the shock front is a sharp jump in temperature, taken along the collapsed characteristics. This differs from the MD results where the temperature distribution is smeared on the shock front. 

\begin{figure}[htb]
	\captionsetup{type=figure}
	\subfloat[Evolution of density (MD)]{\includegraphics[trim=.5cm .5cm 0cm 1cm, clip=true, width=0.51\linewidth]{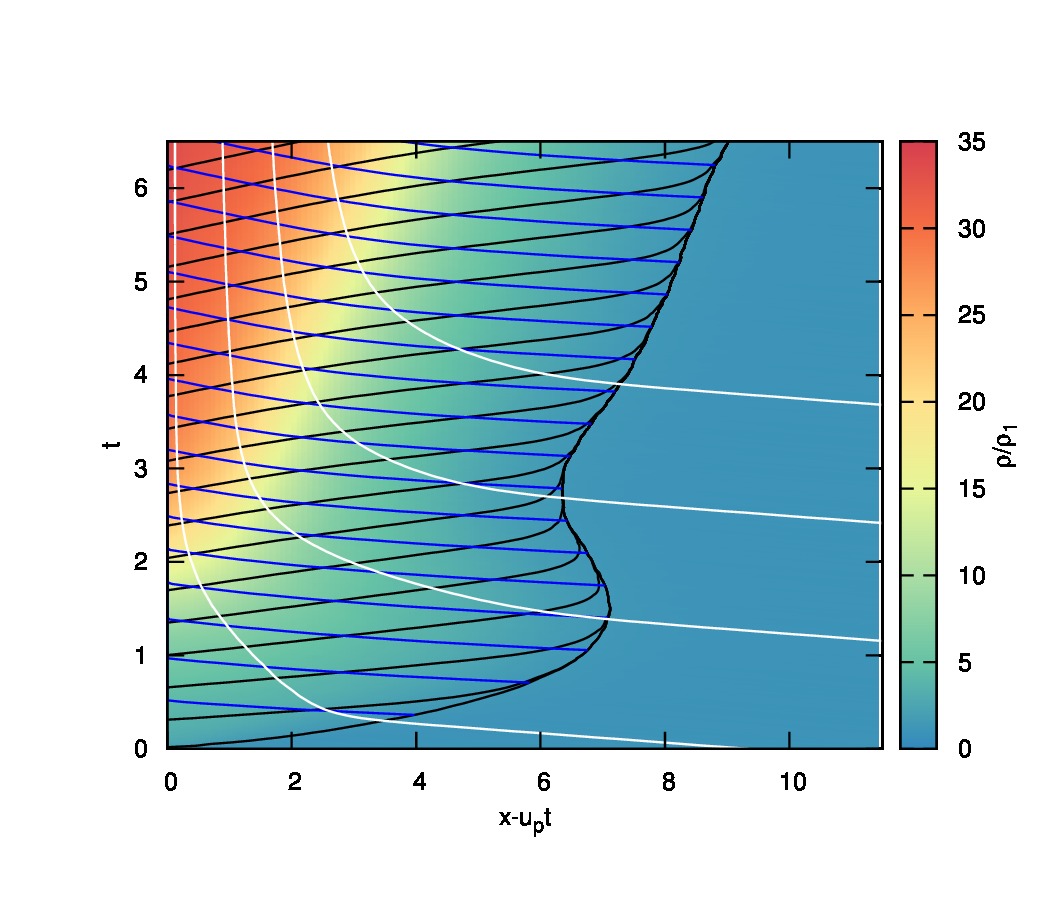}}~
	\subfloat[Evolution of pressure (MD)]{\includegraphics[trim=.5cm .5cm 0cm 1cm, clip=true, width=0.51\linewidth]{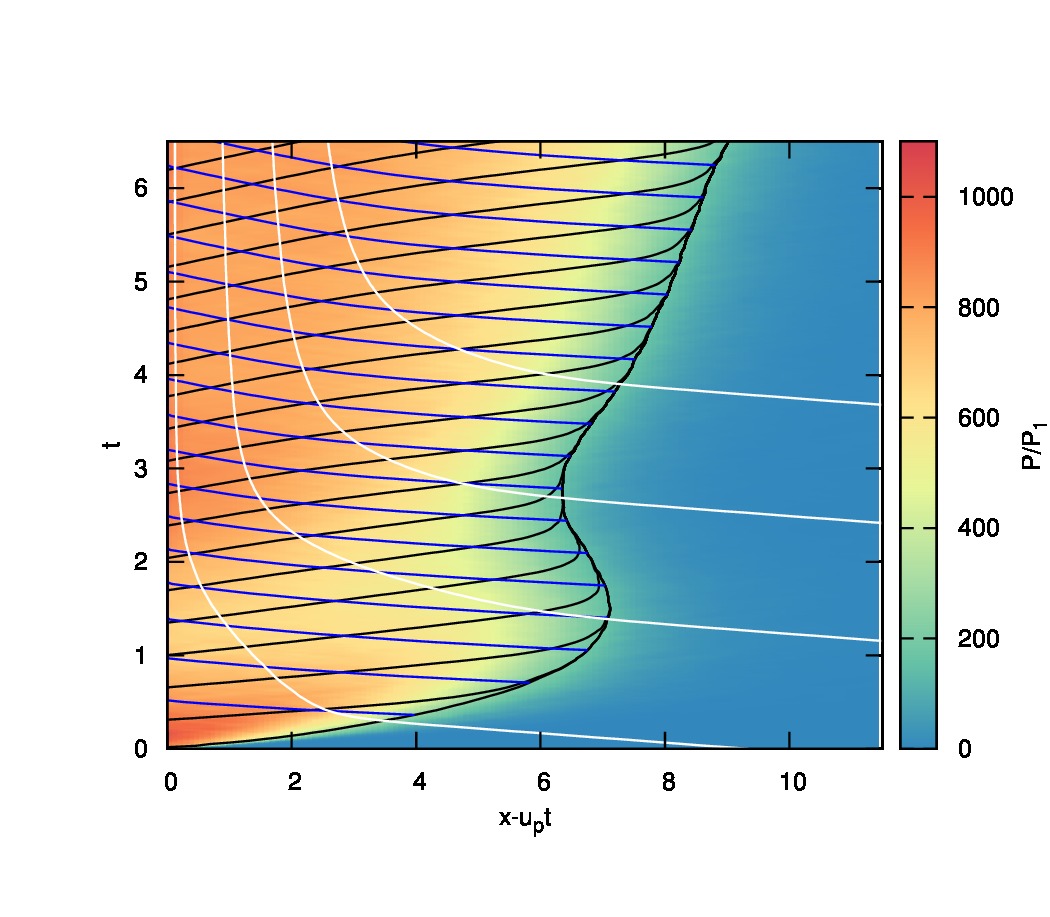}}
	\caption{Evolution of density and pressure on an $x$ vs.~$t$ plane, in the piston frame of reference, from MD model, for $u_p=20$, $u^*=10$ and $\varepsilon=0.95$. Also shown are select particle paths (white), forward (black) and backward (blue) running characteristics.}
	\label{fig:xt_pframe_MD}
\end{figure}

\begin{figure}[htb]
	
	\captionsetup{type=figure}
	\captionsetup[subfigure]{labelformat=empty, oneside,margin={0cm,0cm}, type=figure} 
	
	\includegraphics[width=1\linewidth]{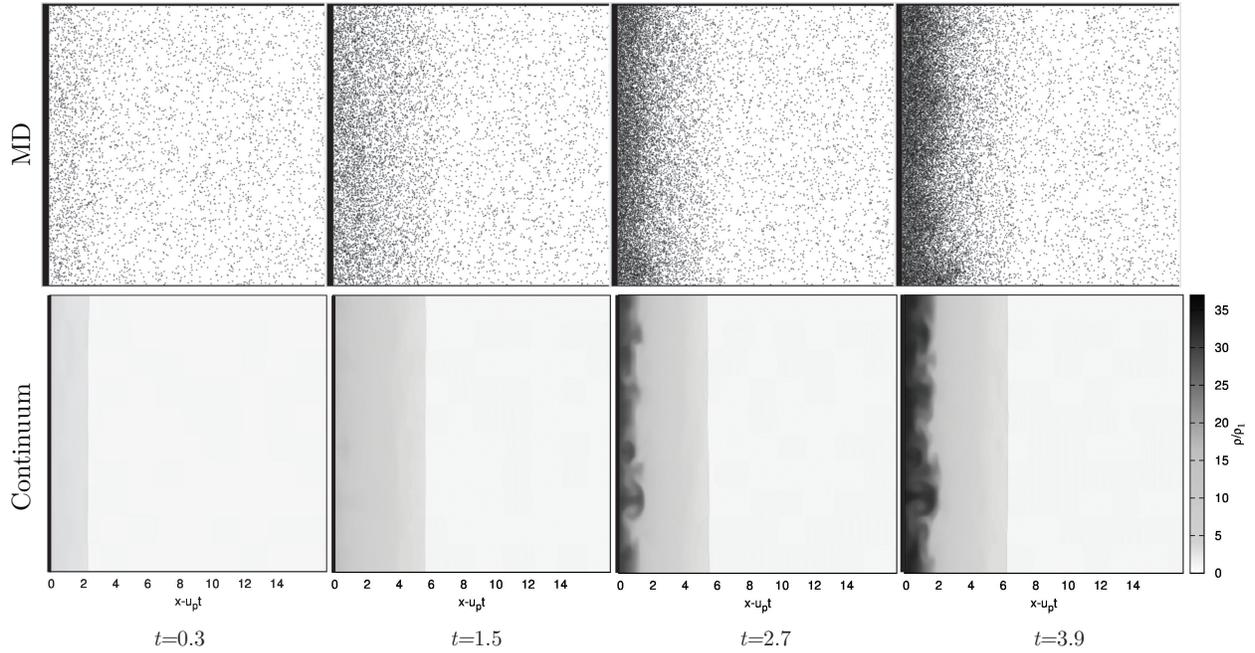}
	\caption{Comparison of the evolution of shock morphology from MD (top) and continuum (bottom) models for $u_p=20$, $u^*=10$ and $\varepsilon=0.95$. The MD solution shows the position of each particle, while the continuum solution shows the distribution of density ratio $\rho/\rho_1$. }
	\label{fig:compareMDHD2D}
\end{figure}
Initially, the solutions have a strong jump in temperature, generated by a faster shock wave. The shocks decelerate, eventually pulling back towards the piston, thus recovering what is observed for purely inelastic media \cite{Kamenetsky_etal2000}. The shock then accelerates and tends to the developed structure. The shock obtained from the continuum model does not extend as far ahead of the piston, which may be due to the under-approximation of pressure discussed earlier. Since the evolutions are similar for both models, we will henceforth look at the evolutions obtained via MD.

To further investigate the stages of the evolution, including the pullback phenomenon, we look at the interaction of the characteristics with the evolution of temperature, density and pressure. The evolutions of density and pressure are shown in Figure \ref{fig:xt_pframe_MD}, along with select particle paths, $C^+$ and $C^-$ characteristics.

The first stages in the evolution are related to the decay of shock front. This can be seen by the $C^+$ waves traversing through particle paths that are undergoing a strong temperature decay, which is causing the shock to slow down. When the shock stops and begins pulling towards the piston, a $C^-$ wave extends to the piston. At the point where this characteristic meets the piston, the equilibrium zone begins. This is seen by the high density, low temperature region, beginning at $t\approx2$. Once $C^+$ waves extend from this equilibrium zone, and meet the shock front, the pullback ends and the shock begins extending from this piston. As the equilibrium zone develops, and $C^+$ characteristics meet the front, a constant front velocity is attained.

Figure \ref{fig:compareMDHD2D} contains snapshots of the evolution of shock morphology for this case, generated from both models. Initially, there are no instabilities visible in both solutions, with a planar shock extending away from the piston. This is seen at $t=0.3$, and up to $t=1.5$, which is the point where the shock front stops propagating ahead of the piston. For later times, high density non-homogeneities appear at the piston face. This is seen at $t=2.7$, confirming that these instabilities occur between $t=1.5$ and $t=2.7$. Comparing with the evolution of pressure, this range in time is when the early particle paths undergo a re-pressurization event on route to attaining an equilibrium state.  Interestingly, this re-pressurization coincides with the time that rear running pressure waves from the pulling back shock front meet the piston, suggesting a possible relationship between these gas dynamic events and the ensuing instability. Once the shock evolution enters to the developed stage, the clusters begin growing from the piston, as demonstrated by the snapshot at $t=3.9$.

\vspace{-1pt}
\subsection{Evolution for varying $u_p/u^*$}
\vspace{-1pt}
\noindent The example shown above is for strong shocks which have been shown to become unstable. Previous results have shown that for decreasing $u_p$ the shocks become stable in the MD simulations for $u_p/u^* \lesssim 1$. In this section we investigate how decreasing $u_p$ affects the evolution of the shock waves, which may have direct bearing on the stability's dependence on $u_p/u^*$.

\begin{figure}[htbp]
\captionsetup{type=figure}
\includegraphics[width=1.0\linewidth]{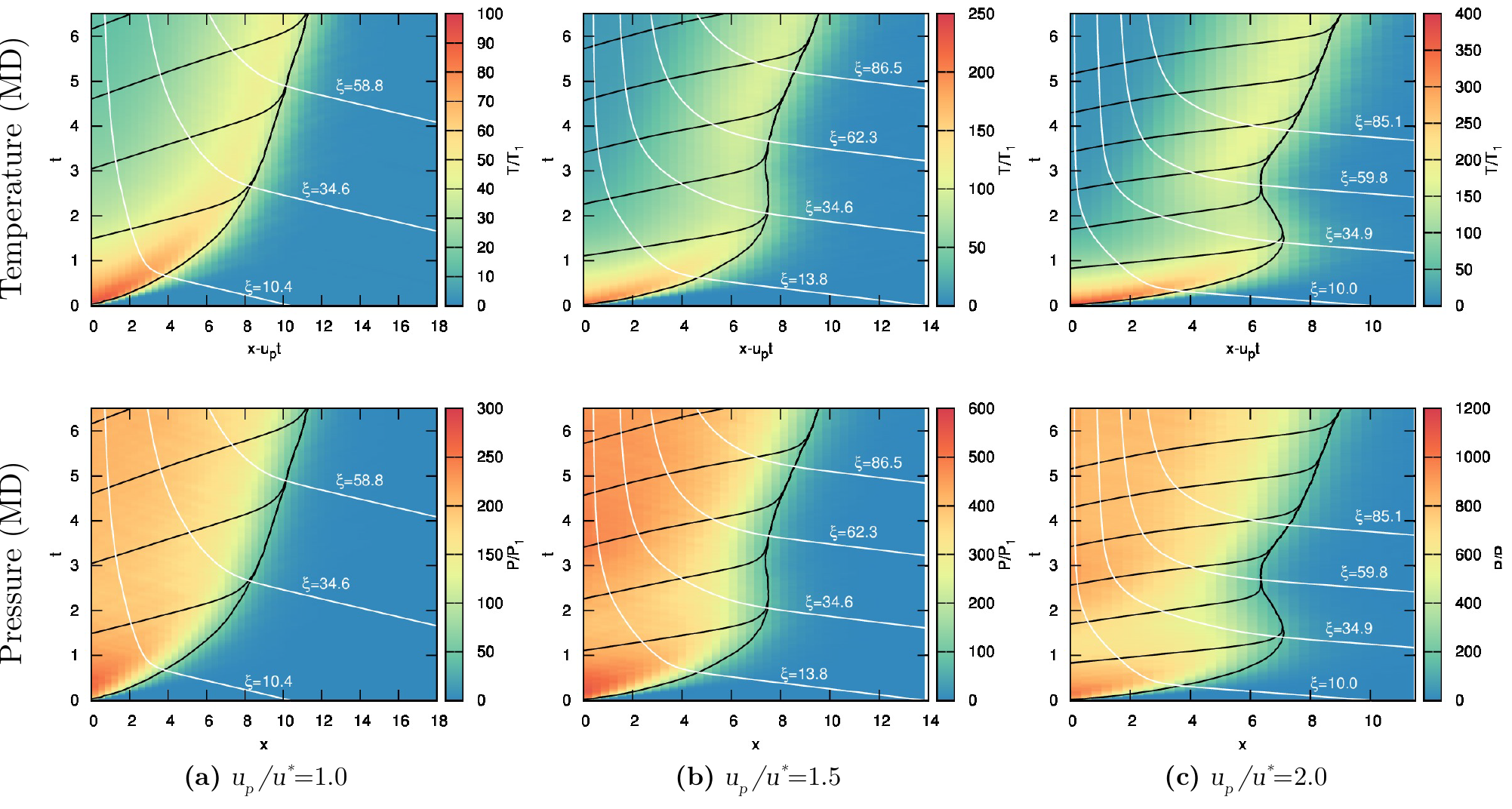}\\
\label{fig:compareXT}
\caption{Comparison of the shock evolution for temperature (top) and pressure (bottom) obtained from MD for varying $u_p/u^*$, with $u^*=10$, $\varepsilon=0.95$ and $\eta_1=0.012$. Selected particle paths (white) and forward running characteristics (black) are shown, where $\xi=x(t=0)$.}
\end{figure}

Figure \ref{fig:compareXT} compares the temperature and pressure evolutions obtained from the MD models for $u_p/u^*=2.0$, 1.5, and 1.0, with $\varepsilon=0.95$. Figure \ref{fig:compareXT}(a) shows results for $u_p/u^*=1.0$, which shows no pullback. Instead, the strong initial shock wave is followed by a gradual decay of the shock velocity. In this case, the early particle paths do not experience a re-pressurization along the piston face.
Increasing $u_p/u^*$, seen in Figure \ref{fig:compareXT}(b) for $u_p/u^*=1.5$, there is a stall in the shock front from the increasing rate of decay.
For this case there is also a re-pressurization event, similar to that seen for $u_p/u^*=2.0$ (Figure \ref{fig:compareXT}(c)), which generates pullback. These results suggest that the rate of decay directly influences this pullback and re-pressurization, in conjunction with rate of communication from the family of characteristics.

We take a Lagrangian approach to quantify the rate of decay, tracking the decay of particles once they are shocked. The evolutions in Figure \ref{fig:compareXT} shows selected particle paths, that originate from $\xi=x(t=0)$. Once shocked, these paths tend to the piston velocity (parallel to the $y$-axis). By tracking the temperature along each of these particle paths once they cross the shock front, we are able to observe how the decay varies for each element. This is done by fitting the shocked temperature evolution to an exponential decay, $T/T_1=A\exp [{-t/\tau_R}]+B$, where $\tau_R$ is the relaxation time constant. 

\begin{SCfigure}[][btp]
\centering
\includegraphics[width=.5\linewidth]{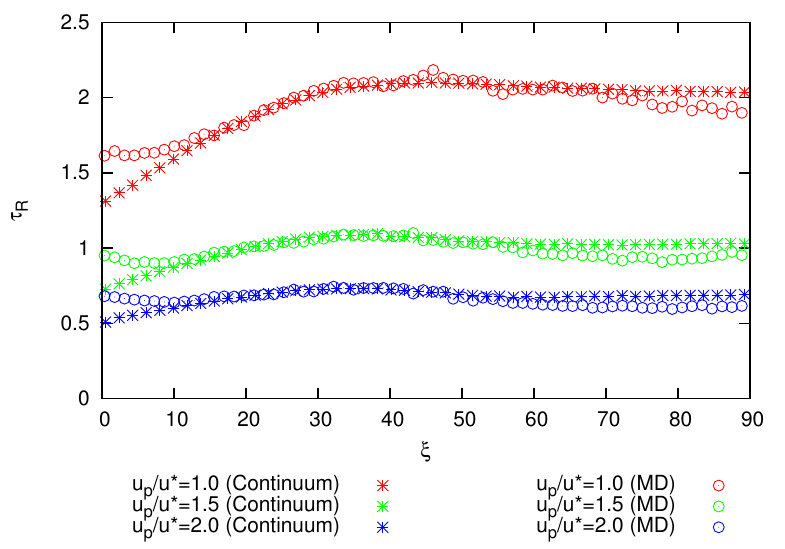} 
\caption{Comparison of relaxation time scales $\tau_R$ obtained for each shocked particle path, where $\xi$~$=$~$x(t=0)$, for varying $u_p/u^*$ and $\varepsilon=0.95$.}
\label{fig:Botheps95trelax_all}
\end{SCfigure}

Figure \ref{fig:Botheps95trelax_all} shows the time constants obtained for varying $u_p/u^*$ with $\varepsilon=0.95$. Results show that $\tau_R$ decreases with increasing $u_p/u^*$, increasing the rate of relaxation and decaying faster once shocked. The higher relaxation rate for the continuum model at early times explains why the shock does not extend as far ahead of the piston, initially slowing the shock front at a faster rate. 
Eventually, $\tau_R$ tends to a constant for each $u_p/u^*$, corresponding to the particle paths crossing the shock front during the developed stage. Note that later discrepancies in the MD model may be due to the finite domain length, since particle paths are restricted from fully entering the equilibrium region. These results show timescales comparable to the time it takes for equilibrium zone to be communicated to the shock front via $C^+$ waves.

\vspace{-1pt}
\subsection{Evolution for varying $\varepsilon$}
\begin{figure}[htb]
\centering
\captionsetup{type=figure}

\subfloat[Shock evolutions for varying $\varepsilon$]{\includegraphics[width=0.48\linewidth]{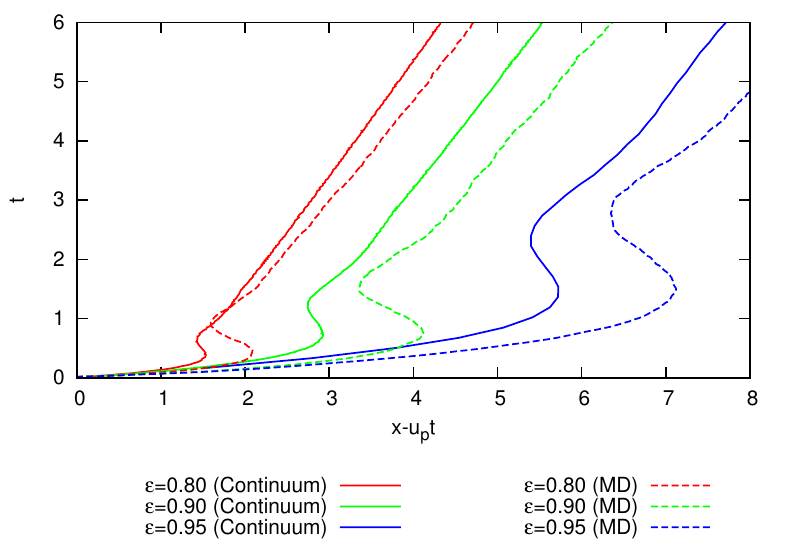}}\hspace{20pt}

\subfloat[Relaxation time scales for varying $\varepsilon$] {\includegraphics[width=.47\linewidth]{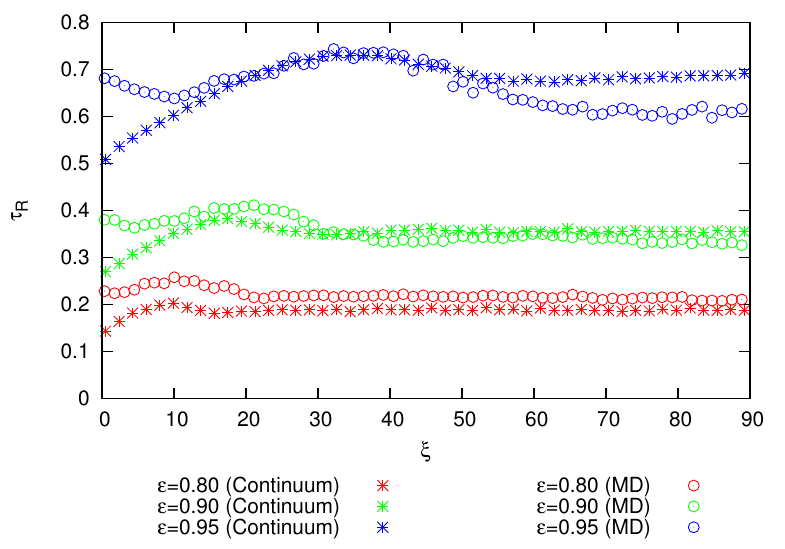}}
\caption{Evolution of shock front (left) for varying $\varepsilon$, with $u^*=10$ and $u_p=20$, along with relaxation time scales $\tau_R$ obtained for each shocked particle path (right), where $\xi=x(t=0)$.}
\label{fig:Bothdiffepsrelax_all}
\end{figure}

\vspace{-1pt}
\noindent We investigate how the shock evolution varies with $\varepsilon$, while maintaining $u_p/u^*=2.0$. Figure \ref{fig:Bothdiffepsrelax_all}(a) shows the shock front evolution for $\varepsilon=0.95$, 0.90, and 0.80, obtained from both models. From these figures, we see that decreasing $\varepsilon$ generates a more rapid decay of the shock front. This is shown by the shock pulling towards the piston after a shorter time. These shocks are also closer to the piston, representing a more tightly packed relaxing region. Although shocks develop faster with decreasing $\varepsilon$, all shocks tend to the same developed velocity. 

As shown previously for $\varepsilon=0.95$, the shocks from the continuum model do not extend as far ahead of the piston as the MD solution, which is now seen for all values of $\varepsilon$. 

The values of $\tau_R$ are obtained for varying $\varepsilon$, with the results shown in Figure \ref{fig:Bothdiffepsrelax_all}(b). As expected, $\tau_R$ decreases with decreasing $\varepsilon$, representing a more rapid decay. The initial values of $\tau_R$ are lower for the continuum model, agreeing with the pullback occurring sooner for this model. For decreasing $\varepsilon$, discrepancies between the two models are more evident. This may be due to a combination of the equation of state that was used in the continuum model, and an increasing role that viscosity has for decreasing $\varepsilon$, with our continuum model unable to capture that.

\vspace{-1pt}

\section{Conclusion}
\vspace{-1pt}
\noindent In this study, we investigate the role that an activation threshold for inelastic collisions has on the evolution of piston driven shock waves through a granular gas. This is done at the discrete and continuum levels. The evolution is found to be dependent on the coefficient of restitution $\varepsilon$ and on the ratio of piston velocity to impact velocity threshold $u_p/u^*$.  For weaker piston velocities, where fewer collisions are activated, the evolution differs from that seen in purely inelastic media. For stronger shocks, the behaviour is similar to that seen in purely inelastic media. This is characterized by a short period where the shock front pulls back towards the piston. The onset of instability at the piston face is seen to occur at approximately the same time that the pullback is communicated to the piston via backwards running pressure waves. These results may shed light on the role early stages of development play on the stability of shock waves through granular media, which have been observed experimentally.

\vspace{-1pt}
\begin{theacknowledgments}
\vspace{-1pt}
\noindent The authors would like to thank Sam Falle (University of Leeds) for support in using the \textit{MG} software. NS is a recipient of the Alexander Graham Bell Canada Graduate Scholarship (NSERC). 
\end{theacknowledgments}
\vspace{-1pt}
\bibliographystyle{aipproc}   

\bibliography{references}

\begin{thebibliography}{16}
\expandafter\ifx\csname natexlab\endcsname\relax\def\natexlab#1{#1}\fi
\providecommand{\enquote}[1]{``#1''}
\expandafter\ifx\csname url\endcsname\relax
  \def\url#1{\texttt{#1}}\fi
\expandafter\ifx\csname urlprefix\endcsname\relax\def\urlprefix{URL }\fi
\providecommand{\eprint}[2][]{\url{#2}}

\bibitem[Frost et~al.(2012)]{Frost2012}
D.~L. Frost, Y.~Gregoire, O.~Petel, S.~Goroshin, and F.~Zhang, \emph{Phys.
  Fluids} \textbf{29}, 091109 (2012).

\bibitem[Rodriguez et~al.(2013)]{Rodriguez2013}
V.~Rodriguez, R.~Saurel, G.~Jourdan, and L.~Houas, \emph{Phys. Rev. E}
  \textbf{88}, 063011 (2013).

\bibitem[Boudet and Kellay(2013)]{Boudet2013}
J.~F. Boudet, and H.~Kellay, \emph{Phys. Rev. E} \textbf{87} (2013).

\bibitem[Bizon et~al.(1998)]{Bizon1998}
C.~Bizon, M.~D. Shattuck, J.~B. Swift, W.~D. McCormick, and H.~L. Swinney,
  \emph{Phys. Rev. Lett.} \textbf{80}, 57--60 (1998).

\bibitem[Carrillo et~al.(2008)]{Carrilo2008}
J.~A. Carrillo, T.~P\"{o}schel, and C.~Saluena, \emph{J. Fluid Mech.}
  \textbf{591}, 199--144 (2008).

\bibitem[Radulescu and Sirmas(2012)]{RadulescuSirmas2011}
M.~Radulescu, and N.~Sirmas, \emph{AIP Conf. Proc.} \textbf{1426}, 1631--1634
  (2012).

\bibitem[Sirmas et~al.(2014)]{Sirmasetal2013}
N.~Sirmas, S.~Falle, and M.~Radulescu, \emph{J. Phys.: Conf. Ser.} \textbf{500}
  (2014).

\bibitem[Goldshtein et~al.(1996)]{Goldshteinetalch31996}
A.~Goldshtein, M.~Shapiro, and C.~Gutfinger, \emph{J. Fluid Mech.} \textbf{316}
  (1996).

\bibitem[Kamenetsky et~al.(2000)]{Kamenetsky_etal2000}
V.~Kamenetsky, A.~Goldshtein, M.~Shapiro, and D.~Degani, \emph{Phys. Fluids}
  \textbf{12}, 3036--3049 (2000).

\bibitem[Sirmas(2013)]{Sirmas}
N.~Sirmas, {M.A.Sc.} thesis {U}niversity of {O}ttawa (2013).

\bibitem[P\"{o}schel et~al.(2003)]{Poscheletal2003}
T.~P\"{o}schel, N.~V. Brilliantiov, and T.~Schwager, \emph{Physica A}
  \textbf{325}, 274--283 (2003).

\bibitem[P\"{o}schel and Schwager(2005)]{Poschel&Schwager2005}
T.~P\"{o}schel, and T.~Schwager, \emph{Computational granular dynamics: models
  and algorithms}, Springer-Verlag, Berlin New York, 2005.

\bibitem[Torquato(1995)]{Torquato1995}
S.~Torquato, \emph{Phys. Rev. E} \textbf{51}, 3170--3182 (1995).

\bibitem[Brilliantov and P\"{o}schel(2004)]{Brilliantov&Poschel2004}
N.~Brilliantov, and T.~P\"{o}schel, \emph{Kinetic theory of granular gases},
  Oxford University Press, Oxford, 2004.

\bibitem[Landau and Lifshitz(1987)]{Landau&Lifshitz1987}
L.~D. Landau, and E.~M. Lifshitz, \emph{Fluid mechanics},
  Butterworth-Heinemann, 1987, 2nd. edn.

\bibitem[Sirmas et~al.(2012)]{Sirmasetal2012}
N.~Sirmas, M.~Tudorache, J.~Barahona, and M.~I. Radulescu, \emph{Shock Waves}
  \textbf{22}, 237--247 (2012).

\end{thebibliography}

\IfFileExists{\jobname.bbl}{}
{\typeout{}
	\typeout{******************************************}
	\typeout{** Please run "bibtex \jobname" to optain}
	\typeout{** the bibliography and then re-run LaTeX}
	\typeout{** twice to fix the references!}
	\typeout{******************************************}
	\typeout{}
}

\end{document}